\documentclass[prd,aps,showpacs,nofootinbib,preprint,eqsecnum]{revtex4}
\usepackage{graphicx,color,amsmath,amsxtra}
\usepackage{epsf}
\usepackage{amssymb}
\usepackage{enumerate}
\usepackage{hhline}
\usepackage{array}
\usepackage{tabularx}

\newcommand{\bear}{\begin{array}}  \newcommand{\eear}{\end{array}}
\newcommand{\bea}{\begin{eqnarray}}  \newcommand{\eea}{\end{eqnarray}}
\newcommand{\beq}{\begin{equation}}  \newcommand{\eeq}{\end{equation}}
\newcommand{\bef}{\begin{figure}}  \newcommand{\eef}{\end{figure}}
\newcommand{\bec}{\begin{center}}  \newcommand{\eec}{\end{center}}

\newcommand{\Eqn}[1]{&\hspace{-0.2em}#1\hspace{-0.2em}&}

\def\be{\begin{equation}}
\def\ee{\end{equation}}
\def\bea{\begin{eqnarray}}
\def\eea{\end{eqnarray}}
\def\beq{\begin{eqnarray}}
\def\eeq{\end{eqnarray}}

\def\be{\begin{equation}}
\def\ee{\end{equation}}
\def\bea{\begin{eqnarray}}
\def\eea{\end{eqnarray}}
\def\beq{\begin{eqnarray}}
\def\eeq{\end{eqnarray}}

\begin{document}

\title{Thermodynamics in $f(R)$ gravity in the Palatini formalism
}

\author{Kazuharu Bamba\footnote{E-mail address: 
bamba@phys.nthu.edu.tw} and 
Chao-Qiang Geng\footnote{E-mail address: geng@phys.nthu.edu.tw} 
}
\affiliation{
Department of Physics, National Tsing Hua University, Hsinchu, Taiwan 300 
}


\begin{abstract}

We investigate thermodynamics of the apparent horizon in $f(R)$ gravity 
in the Palatini formalism with non-equilibrium and equilibrium descriptions. 
We demonstrate that it is more transparent to understand the horizon entropy 
in the equilibrium framework than that in the non-equilibrium one. 
Furthermore, we show that the second law of thermodynamics can be explicitly 
verified in both phantom and non-phantom phases for the same temperature of 
the universe outside and inside the apparent horizon. 

\end{abstract}


\maketitle

\section{Introduction}

There are two representative approaches to account for the current accelerated 
expansion~\cite{WMAP, SN1} of the universe~\cite{Peebles:2002gy, Sahni:2005ct, 
Padmanabhan:2002ji, Copeland:2006wr, D-and-M, Nojiri:2006ri, rv-2, 
Sotiriou:2008rp, Lobo:2008sg, Capozziello, Silvestri:2009hh, Sami:2009jx, 
Cai:2009zp}. 
One is the introduction of ``dark energy'' in the framework of general 
relativity. 
The other is the investigation of a modified gravitational theory, such as 
$f(R)$ gravity~\cite{f(R)-Capo, Carroll:2003wy, NO-i-s, Chiba:2003ir}. 
In this study, we concentrate on thermodynamics 
in the modified gravitational theory. 
The fundamental connection between gravitation and thermodynamics 
has been suggested by the discovery of black hole 
thermodynamics~\cite{Bardeen:1973gs} with black hole 
entropy~\cite{Bekenstein:1973ur} and Hawking 
temperature~\cite{Hawking:1974sw}. 
Its application to the cosmological event horizon of de Sitter 
space has been explored~\cite{Gibbons:1977mu} (for recent reviews, 
see~\cite{Rev-T-G-Pad}). 
It was shown that the Einstein equation can be derived from the 
proportionality of the entropy to the horizon area together with 
the Clausius relation in thermodynamics~\cite{Jacobson:1995ab}. 
This consequence has been applied to various cosmological 
settings~\cite{Frolov:2002va, Danielsson:2004xw, Bousso:2004tv, Cai:2005ra, 
Akbar:2008vc}. 
In Ref.~\cite{Cai:2005ra}, it was shown that if the entropy of the apparent 
horizon in the Friedmann-Lema\^{i}tre-Robertson-Walker (FLRW) spacetime is 
proportional to the apparent horizon area, Friedmann equations follow from the 
first law of thermodynamics. 

Thermodynamics in dynamical black holes~\cite{Hayward:1993wb, Hayward:1997jp, 
Hayward:1998ee} and spherically symmetric 
spacetimes~\cite{Padmanabhan:2002sha} has also been developed. 
The first law of the ordinary equilibrium thermodynamics has been 
discussed in $f(R)$ gravity, scalar-tensor theories, the Gauss-Bonnet gravity 
and more general Lovelock gravity~\cite{Akbar:2006er, Akbar:2006kj, TD-LL-Pad, 
Cai:2008mh, Cai:2009de}, while the second law has been studied in the 
accelerating universe, $f(R)$ gravity, the Gauss-Bonnet gravity and the 
Lovelock gravity in 
Refs.~\cite{Zhou:2007pz, MohseniSadjadi:2007zq, Akbar:2008vz}, respectively. 
Thermodynamics in braneworld scenario~\cite{Braneworld-scenario}, 
loop quantum cosmology~\cite{Cai:2008ys, LQC-Zhu} 
and Ho\v{r}ava-Lifshitz gravity~\cite{HL-gravity} 
as well as its properties of dark energy~\cite{Gong:2006ma, Gong:2006sn, 
Saridakis:2009uu} 
and those in cosmological scenarios with interactions between dark energy and 
dark matter~\cite{Jamil:2009eb, Interacting-scenario-S} have been also 
examined. 
Incidentally, the horizon entropy in the four-dimensional 
modified gravity~\cite{Wang:2005bi} and the quantum logarithmic correction to 
the expression of the horizon entropy in cosmological 
contexts~\cite{Cai:2008ys, Lidsey:2008zq, Zhu:2008cg, Cai:2009ua} have been 
investigated. 
In the framework of general relativity, 
the considerations on the sign of entropy~\cite{Brevik:2004sd, Bilic:2008zk} 
and its evolution~\cite{Buchmuller:2006em} in the phantom phase have been 
performed. 
Moreover, the conditions~\cite{Briscese:2007cd} that the black hole entropy 
can be positive in the $f(R)$ gravity 
models~\cite{Hu:2007nk, Starobinsky:2007hu, Appleby:2007vb} with passing 
the solar-system tests and cosmological bounds 
and the evolution of the black hole entropy in the 
ghost condensate scenario~\cite{Mukohyama:2009um} 
have been studied. 

It was pointed out~\cite{Eling:2006aw} that in $f(R)$ gravity, 
a non-equilibrium thermodynamic treatment should be required 
in order to derive the corresponding gravitational field equation by using the 
procedure in Ref.~\cite{Jacobson:1995ab}. 
This point was reanalyzed in $f(R)$ gravity~\cite{Akbar:2006mq} as well as 
scalar-tensor theories~\cite{Cai:2006rs} and elaborated by considering 
the role of gravitational dissipation~\cite{Chirco:2009dc}. 
By taking into account the non-equilibrium thermodynamic treatment, 
the studies on the first~\cite{Wu:2007se} and second~\cite{Wu:2008ir} laws of 
thermodynamics on the apparent horizon in generalized theories of gravitation 
and the investigations on thermodynamics~\cite{Bamba:2009ay} in a 
$f(R)$ gravity model~\cite{Bamba:2008hq} with realizing a crossing of the 
phantom divide from the non-phantom (quintessence) phase to the phantom one 
have been performed. 
The non-equilibrium correction~\cite{Eling:2006aw} has been reinterpreted 
through the introduction of a mass-like function~\cite{Gong:2007md} and by 
other approaches~\cite{Eling:2008af, Elizalde:2008pv, Wu:2009wp}. 
The formulation in Ref.~\cite{Jacobson:1995ab} was also extended to 
the more general extended gravity theory~\cite{Brustein:2009hy, Bamba:2009gq} 
(for related discussions, see~\cite{Parikh:2009qs}). 

Recently, it has been shown that it is possible to obtain a picture of 
equilibrium thermodynamics on the apparent horizon in the FLRW background 
for modified gravity theories with the Lagrangian density $f(R, \phi, X)$ 
(including $f(R)$ gravity and scalar-tensor theories), 
where $X=-\left(1/2\right) g^{\mu\nu} {\nabla}_{\mu}\phi {\nabla}_{\nu}\phi$
is the kinetic term of a scalar field $\phi$ (${\nabla}_{\mu}$ is the 
covariant derivative operator associated with the metric tensor 
$g_{\mu \nu}$), due to a suitable redefinition of an energy momentum tensor of 
the ``dark'' component that respects a local energy 
conservation~\cite{Bamba:2009id}. 

The previous studies on thermodynamics in $f(R)$ gravity have been executed 
in the standard metric formalism, in which the connection is defined in terms 
of the metric. 
There exist other formalisms to derive the gravitational field equation from 
the action, such as the Palatini formalism, in which the connection and the 
metric are assumed to be independent variables (for more detailed explanations 
on the Palatini formalism, see~\cite{Sotiriou:2008rp}). 
As a result, the gravitational field equation of $f(R)$ gravity in 
the Palatini formalism is different from that in the metric one. 
Furthermore, $f(R)$ gravity~\cite{Palatini-f(R)-1, Olmo:2005zr, Olmo:2005jd, 
Olmo:2008nf, Olmo:2008ye} 
and more general gravitational theories~\cite{Allemandi:2004wn, Li:2007xw} 
in the Palatini formalism have been examined. 
Bouncing cosmologies in the Palatini $f(R)$ 
gravity~\cite{Barragan:2009sq} 
and the dynamical aspects of generalized Palatini theories of 
gravity~\cite{Olmo:2009xy} have been also investigated. 

It should be mentioned that the Palatini $f(R)$ gravity 
(with the exception of those with very heavily suppressed correction which 
would only become significant in the very far UV) is in conflict with 
Solar system tests~\cite{Olmo:2005jd} 
and particle physics~\cite{Flanagan:2003rb, Iglesias:2007nv}. 
Moreover, it is plagued by surface singularities in stars and spherical matter 
configurations~\cite{B-S-M} 
and it does not necessarily have a well posed Cauchy problem in the presence 
of matter~\cite{Sotiriou:2008rp}. 
However, it should be cautioned that some of the arguments may not be generic. 
For example, most of criticism were based on $f(R) = R - \beta/R^n$, where 
$\beta$ and $n$ are constants, 
while the against of the Palatini $f(R)$ gravity based on 
particle physics may be premature as the Higgs sector is not well understood 
yet. 

On the other hand,
although there have been many studies on thermodynamics in $f(R)$ gravity
in the literature, an equilibrium thermodynamics description of $f(R)$ gravity 
in the Palatini formalism has not been done yet. 
In particular,
it is interesting to ask whether
the non-equilibrium and 
equilibrium descriptions of thermodynamics can be given in the Palatini 
formalism as the metric one~\cite{Bamba:2009id}. 
For the interest of the formalism, 
in this paper we explore both non-equilibrium and 
equilibrium descriptions of thermodynamics 
in the Palatini formalism of $f(R)$ gravity. 
We also show that  the second law of 
thermodynamics in $f(R)$ gravity can be explicitly verified in the phantom 
phase as well as the non-phantom (quintessence) one
if the temperature of the universe inside the horizon is 
equal to that of the apparent horizon. 
We use units of $k_\mathrm{B} = c = \hbar = 1$ and denote the
gravitational constant $8 \pi G$ by 
${\kappa}^2 \equiv 8\pi/{M_{\mathrm{Pl}}}^2$ 
with the Planck mass of $M_{\mathrm{Pl}} = G^{-1/2} = 1.2 \times 10^{19}$GeV.

The paper is organized as follows. 
In Sec.\ II, we explain $f(R)$ gravity in the Palatini formalism and 
study the first and second laws of thermodynamics of the 
apparent horizon in $f(R)$ gravity under 
a non-equilibrium description of thermodynamics. 
In Sec.\ III, we illustrate an equilibrium description of thermodynamics 
by redefining the energy density and pressure of dark components 
in $f(R)$ gravity in the Palatini formalism. 
Finally, conclusions are given in Sec.\ IV.

\section{Non-equilibrium description of thermodynamics in $f(R)$ gravity 
in the Palatini formalism}

\subsection{$f(R)$ gravity in the Palatini formalism}

The action of $f(R)$ gravity in the Palatini formalism with matter is written 
as
\begin{eqnarray}
I = \int d^4 x \sqrt{-g} \left[ \frac{f(\mathcal{R})}{2\kappa^2} +
{\mathcal{L}}_{\mathrm{matter}} \right]\,,
\label{eq:2.1}
\end{eqnarray}
where $g$ is the determinant of the metric tensor $g_{\mu\nu}$ and
${\mathcal{L}}_{\mathrm{matter}}$ is the matter Lagrangian. 
In the Palatini formalism, the connection 
and the metric tensor $g_{\mu\nu}$ are treated as independent variables. 
We denote the Ricci tensor constructed with this independent connection 
as ${\mathcal{R}}_{\mu \nu}$ and the corresponding Ricci scalar as 
$\mathcal{R} = g^{\mu\nu} {\mathcal{R}}_{\mu \nu}$. 
In general, ${\mathcal{R}}_{\mu \nu}$ is different from the Ricci tensor 
$R_{\mu \nu}$ constructed with the Levi-Civita connection of the metric. 
Here, the Ricci scalar $R$ is given by $R=g^{\mu\nu} R_{\mu \nu}$ and 
$f(\mathcal{R})$ is an arbitrary function of $\mathcal{R}$. 

Taking the variation of the action in Eq.~(\ref{eq:2.1}) with respect to 
$g_{\mu\nu}$, one obtains~\cite{Sotiriou:2008rp} 
\begin{eqnarray}
F(\mathcal{R}){\mathcal{R}}_{\mu \nu}
-\frac{1}{2}f(\mathcal{R})g_{\mu\nu} = 
\kappa^2 T^{(\mathrm{matter})}_{\mu \nu}\,, 
\label{eq:2.2}
\end{eqnarray}
where 
$F(\mathcal{R}) \equiv d f(\mathcal{R})/d \mathcal{R}$. 
Here, 
$T^{(\mathrm{matter})}_{\mu \nu}$ 
is the contribution to the energy-momentum tensor from all 
perfect fluids of ordinary matter (radiation and non-relativistic matter) 
with $\rho_{\mathrm{f}}$ and $P_{\mathrm{f}}$ being the energy density and 
pressure of all ordinary matters, respectively. 

Varying the action in Eq.~(\ref{eq:2.1}) with respect to the 
connection and using Eq.~(\ref{eq:2.2}), one finds~\cite{Sotiriou:2008rp} 
\begin{equation}
F G_{\mu\nu} 
= 
\kappa^2 T^{(\mathrm{matter})}_{\mu \nu} 
-\frac{1}{2} g_{\mu \nu} \left( F\mathcal{R} - f \right)
+ \nabla_{\mu}\nabla_{\nu}F -g_{\mu \nu} \Box F 
-\frac{3}{2F} \left[ \nabla_{\mu} F \nabla_{\nu}F 
-\frac{1}{2} g_{\mu \nu} \left( \nabla F \right)^2 \right]\,,
\label{eq:2.3}
\end{equation}
where 
$G_{\mu\nu}=R_{\mu\nu}-\left(1/2\right)g_{\mu\nu}R$ is 
the Einstein tensor, 
${\nabla}_{\mu}$ is the covariant derivative operator associated with 
the Levi-Civita connection of the metric tensor $g_{\mu \nu}$, 
$\Box \equiv g^{\mu \nu} {\nabla}_{\mu} {\nabla}_{\nu}$
is the covariant d'Alembertian for a scalar field, 
and $\left( \nabla F \right)^2 = g^{\mu \nu} \nabla_{\mu}F \nabla_{\nu}F$. 

We assume the four-dimensional Friedmann-Lema\^{i}tre-Robertson-Walker (FLRW) 
spacetime with the metric, 
\begin{eqnarray}
d s^2 = h_{\alpha \beta} d x^{\alpha} d x^{\beta}
+\tilde{r}^2 d \Omega^2\,,
\label{eq:2.4}
\end{eqnarray}
where $\tilde{r}=a(t)r$, $x^0=t$ and $x^1=r$ with the two 
dimensional metric $h_{\alpha \beta}={\rm diag}(-1, a^2(t)/[1-Kr^2])$.
Here, $a(t)$ is the scale factor, $K$ is the cosmic curvature, and 
$d \Omega^2$ is the metric of two-dimensional sphere with unit radius. 
In the FLRW background (\ref{eq:2.4}), from Eq.~(\ref{eq:2.3}) we obtain the 
following gravitational field equations: 
\begin{eqnarray} 
3F \left( H^2+\frac{K}{a^2} \right) 
\Eqn{=}
\kappa^2 \rho_{\mathrm{f}} +\frac{1}{2} \left( F\mathcal{R} - f \right) 
-3H\dot{F} -\frac{3}{4}\frac{\dot{F}^2}{F}\,,
\label{eq:2.5} \\ 
-2F \left( \dot{H}-\frac{K}{a^2} \right) 
\Eqn{=}
\kappa^2 \left( \rho_{\mathrm{f}} + P_{\mathrm{f}} \right)
+\ddot{F}-H\dot{F} -\frac{3}{2}\frac{\dot{F}^2}{F}\,,
\label{eq:2.6}
\end{eqnarray} 
where $H=\dot{a}/a$ is the Hubble parameter and 
the dot denotes the time derivative of $\partial/\partial t$. 
We note that the perfect fluid satisfies the continuity equation 
\begin{eqnarray} 
\dot{\rho}_{\mathrm{f}}+3H\left( \rho_{\mathrm{f}} + P_{\mathrm{f}} \right)
=0\,.
\label{eq:2.7}
\end{eqnarray} 

Equations (\ref{eq:2.5}) and (\ref{eq:2.6}) can be rewritten as
\begin{eqnarray}  
H^2+\frac{K}{a^2} 
\Eqn{=}
\frac{\kappa^2}{3F} \left( \hat{\rho}_{\mathrm{d}}+
\rho_{\mathrm{f}} \right)\,, 
\label{eq:2.8} \\
\dot{H}-\frac{K}{a^2}
\Eqn{=}
-\frac{\kappa^2}{2F} \left( \hat{\rho}_{\mathrm{d}}+\hat{P}_{\mathrm{d}}
+\rho_{\mathrm{f}}+P_{\mathrm{f}} \right)\,,
\label{eq:2.9} 
\end{eqnarray} 
where $\hat{\rho}_{\mathrm{d}}$ and $\hat{P}_{\mathrm{d}}$ are 
the energy density and pressure of ``dark'' components, given by 
\begin{eqnarray}  
\hat{\rho}_{\mathrm{d}} 
\Eqn{\equiv} 
\frac{1}{\kappa^2} \left[ \frac{1}{2} \left( F\mathcal{R} - f \right) 
-3H\dot{F} -\frac{3}{4}\frac{\dot{F}^2}{F} \right]\,, 
\label{eq:2.10} \\
\hat{P}_{\mathrm{d}} 
\Eqn{\equiv} 
\frac{1}{\kappa^2} \left[ -\frac{1}{2} \left( F\mathcal{R} - f \right) 
+\ddot{F}+2H\dot{F} -\frac{3}{4}\frac{\dot{F}^2}{F} \right]\,. 
\label{eq:2.11}
\end{eqnarray} 
Here, a hat denotes quantities in the non-equilibrium 
description of thermodynamics. 
Note that $\hat{\rho}_d$ and $\hat{P}_d$ are originated from the 
energy-momentum tensor $\hat{T}_{\mu \nu}^{(\mathrm{d})}$, defined by 
\begin{equation}
\hat{T}_{\mu \nu}^{(\mathrm{d})} \equiv 
\frac{1}{\kappa^2} \left\{ 
-\frac{1}{2} g_{\mu \nu} \left( F\mathcal{R} - f \right)
+ \nabla_{\mu}\nabla_{\nu}F -g_{\mu \nu} \Box F 
-\frac{3}{2F} \left[ \nabla_{\mu} F \nabla_{\nu}F 
-\frac{1}{2} g_{\mu \nu} \left( \nabla F \right)^2 \right]
\right\}\,, 
\label{eq:2.12}
\end{equation}
where the Einstein equation is given by 
\begin{eqnarray} 
G_{\mu \nu}=\frac{\kappa^2}{F}
\left( T_{\mu \nu}^{(\rm matter)}+\hat{T}_{\mu \nu}^{(\mathrm{d})} 
\right)\,.
\label{eq:2.13}
\end{eqnarray} 
%

Relation between ${\mathcal{R}}_{\mu \nu}$ and $R_{\mu \nu}$ as well as 
$\mathcal{R}$ and $R$ are given by~\cite{Sotiriou:2008rp}
\begin{eqnarray}
{\mathcal{R}}_{\mu \nu} \Eqn{=} 
R_{\mu \nu} +\frac{3}{2F^2} \nabla_{\mu}F \nabla_{\nu}F 
-\frac{1}{F}\left( \nabla_{\mu}\nabla_{\nu}F 
+\frac{1}{2} g_{\mu \nu} \Box F 
\right)\,, 
\label{eq:2.14} \\ 
\mathcal{R} \Eqn{=} 
R +\frac{3}{2F^2}\left( \nabla F \right)^2 -\frac{3}{F} \Box F\,. 
\label{eq:2.15}
\end{eqnarray} 
In the FLRW background (\ref{eq:2.4}), the scalar curvature in the metric 
formalism is written as 
$R=6 \left( 2H^2 + \dot{H} + K/a^2 \right)$. 
Using this expression and Eq.~(\ref{eq:2.15}), we get 
\begin{eqnarray} 
\mathcal{R} = 6 \left( 2H^2 + \dot{H} + \frac{K}{a^2} \right) + 
\frac{3}{F}\left( -\frac{1}{2}\frac{\dot{F}^2}{F}+\ddot{F}+3H\dot{F} 
\right)\,. 
\label{eq:2.16}
\end{eqnarray}
%

{}From Eqs.~(\ref{eq:2.10}), (\ref{eq:2.11}) and (\ref{eq:2.16}), we find that 
$\hat{\rho}_{\mathrm{d}}$ and $\hat{P}_{\mathrm{d}}$ satisfy the 
following equation: 
\begin{eqnarray}
\dot{\hat{\rho}}_{\mathrm{d}}+3H \left( 
\hat{\rho}_{\mathrm{d}}+\hat{P}_{\mathrm{d}} \right)
=\frac{3}{\kappa^2} \left( H^2+\frac{K}{a^2} \right) \dot{F}\,. 
\label{eq:2.17}
\end{eqnarray} 
In $f(R)$ gravity, since $\dot{F} \neq 0$, the right-hand side (r.h.s.) of 
Eq.~(\ref{eq:2.17}) does not vanish, 
so that the standard continuity equation does not hold.

\subsection{First law of thermodynamics}

We study the thermodynamic property of $f(R)$ gravity. 
The dynamical apparent horizon is determined by the relation
$h^{\alpha \beta} \partial_{\alpha} \tilde{r} \partial_{\beta} \tilde{r}=0$. 
According to the recent type Ia Supernovae data, it is suggested 
that in the accelerating universe the enveloping surface should 
be the apparent horizon rather than the event one from the thermodynamic point 
of view~\cite{Zhou:2007pz}. 
In the FLRW spacetime, the radius $\tilde{r}_A$ of 
the apparent horizon is given by 
\begin{eqnarray}
\tilde{r}_A=\left( H^2+\frac{K}{a^2} \right)^{-1/2}\,.
\label{eq:2.18}
\end{eqnarray} 
Taking the time derivative of this relation, we obtain
\begin{eqnarray}
-\frac{d\tilde{r}_A}{\tilde{r}_A^3}
=\left( \dot{H}-\frac{K}{a^2} \right)H dt\,.
\label{eq:2.19}
\end{eqnarray}
Combing Eqs.~(\ref{eq:2.9}) and (\ref{eq:2.10}), we find
\begin{eqnarray}
\frac{F}{4\pi G} d\tilde{r}_A=\tilde{r}_A^3 H
\left( \hat{\rho}_{\mathrm{d}}+\hat{P}_{\mathrm{d}}+\rho_{\mathrm{f}}
+P_{\mathrm{f}} \right) dt\,.
\label{eq:2.20}
\end{eqnarray}
%

In general relativity, the Bekenstein-Hawking horizon entropy is expressed as 
$S=A/\left(4G\right)$, where $A=4\pi \tilde{r}_A^2$ is the area of the 
apparent horizon~\cite{Bekenstein:1973ur, Bardeen:1973gs, Hawking:1974sw}. 
The Bekenstein-Hawking entropy $S=A/\left(4G\right)$ is a global geometric 
quantity which is proportional to the horizon area $A$ with a constant 
coefficient $1/\left(4G\right)$. 
This quantity is not directly affected by the difference of 
gravitational theories, i.e., the difference of the derivative of the 
Lagrangian density $f$ with respect to $R$, $F$. 
On the other hand, 
in the context of modified gravity theories including $f(R)$ gravity 
a horizon entropy $\hat{S}$ associated with a Noether 
charge has been proposed by Wald~\cite{Wald entropy}. 
The Wald entropy $\hat{S}$ is a local quantity defined 
in terms of quantities on the bifurcate Killing horizon. 
More specifically, it depends on the variation of the Lagrangian 
density of gravitational theories with respect to the Riemann tensor. 
This is equivalent to 
$\hat{S}=A/\left(4G_{\rm eff}\right)$, where $G_{\mathrm{eff}}=G/F$
is the effective gravitational coupling 
in $f(R)$ gravity~\cite{Brustein:2007jj}. 

By using Eq.~(\ref{eq:2.20}) and 
the Wald entropy in $f(R)$ gravity in the Palatini 
formalism~\cite{Vollick:2007fh}
\begin{eqnarray}
\hat{S}=\frac{FA}{4G}\,, 
\label{eq:2.21}
\end{eqnarray} 
we get 
\begin{eqnarray}
\frac{1}{2\pi \tilde{r}_A} d\hat{S}=4\pi \tilde{r}_A^3 H
\left( \hat{\rho}_{\mathrm{d}}+\hat{P}_{\mathrm{d}}+\rho_{\mathrm{f}}
+P_{\mathrm{f}} \right) dt +
\frac{\tilde{r}_A}{2G} dF\,.
\label{eq:2.22}
\end{eqnarray}
%
We note that 
the Wald entropy in $f(R)$ gravity in the metric formalism 
has the same form as in Eq.~(\ref{eq:2.21})~\cite{Wald entropy, 
Jacobson:1993vj, Cognola:2005de}. 

The associated temperature of the apparent horizon has 
the following Hawking temperature $T$ 
determined through the surface gravity $\kappa_{\mathrm{sg}}$: 
\begin{eqnarray}
T \Eqn{=} \frac{|\kappa_{\mathrm{sg}}|}{2\pi}\,, 
\label{eq:2.23} \\
\kappa_{\mathrm{sg}} \Eqn{=} \frac{1}{2\sqrt{-h}} \partial_\alpha 
\left( \sqrt{-h}h^{\alpha\beta} \partial_\beta \tilde{r} \right)\,, 
\label{eq:2.24}
\end{eqnarray}
where $h$ is the determinant of the metric $h_{\alpha\beta}$ and 
$\kappa_{\mathrm{sg}}$ is given by~\cite{Cai:2005ra} 
\begin{eqnarray}
\kappa_{\mathrm{sg}} \Eqn{=} -\frac{1}{\tilde{r}_A}
\left( 1-\frac{\dot{\tilde{r}}_A}{2H\tilde{r}_A} \right)
=-\frac{\tilde{r}_A}{2} \left( 2H^2+\dot{H}
+\frac{K}{a^2} \right) \nonumber \\
\Eqn{=} 
-\frac{2\pi G}{3F} \tilde{r}_A 
\left( \hat{\rho}_T-3\hat{P}_T \right)
\label{eq:2.25}
\end{eqnarray}
with $\hat{\rho}_{\mathrm{t}} \equiv \hat{\rho}_{\mathrm{d}}+
\rho_{\mathrm{f}}$ and $\hat{P}_{\mathrm{t}} \equiv \hat{P}_{\mathrm{d}}+
P_{\mathrm{f}}$, denoted as the total energy density and pressure of the 
universe, respectively. 
It follows from Eq.~(\ref{eq:2.25}) that 
if the total equation of state (EoS) 
$w_{\mathrm{t}}=\hat{P}_{\mathrm{t}}/\hat{\rho}_{\mathrm{t}}$ 
satisfies $w_{\mathrm{t}} \le 1/3$, one has $\kappa_{\mathrm{sg}} \le 0$, 
which is the case for the standard cosmology. 
Using Eqs.~(\ref{eq:2.23}) and (\ref{eq:2.25}), the horizon temperature 
is given by 
\begin{eqnarray}
T=\frac{1}{2\pi \tilde{r}_A}
\left( 1-\frac{\dot{\tilde{r}}_A}{2H\tilde{r}_A} \right)\,.
\label{eq:2.26}
\end{eqnarray}
Multiplying the term $1-\dot{\tilde{r}}_A/(2H\tilde{r}_A)$ for 
Eq.~(\ref{eq:2.22}), we obtain 
\begin{equation}
T d\hat{S} = 4\pi \tilde{r}_A^3 H \left(\hat{\rho}_{\mathrm{d}}+
\hat{P}_{\mathrm{d}}+\rho_{\mathrm{f}}+P_{\mathrm{f}} \right) dt 
-2\pi  \tilde{r}_A^2 \left(\hat{\rho}_{\mathrm{d}}+\hat{P}_{\mathrm{d}}
+\rho_{\mathrm{f}}+P_{\mathrm{f}} \right) d \tilde{r}_A
+\frac{T}{G}\pi \tilde{r}_A^2 dF\,. 
\label{eq:2.27}
\end{equation}

In general relativity, the Misner-Sharp energy~\cite{Misner:1964je, 
Bak:1999hd} is defined as $E=\tilde{r}_A/\left(2G\right)$. 
In $f(R)$ gravity, 
this may be extended to the form~\cite{Gong:2007md, Wu:2007se, Wu:2008ir} 
(for related works, see also Refs.~\cite{Sakai:2001gh, Cai:2009qf})
\begin{eqnarray}
\hat{E}=\frac{\tilde{r}_AF}{2G}\,.
\label{eq:2.28}
\end{eqnarray}
From Eqs.~(\ref{eq:2.18}) and (\ref{eq:2.28}), we find 
\begin{eqnarray}
\hat{E}=\frac{\tilde{r}_A F}{2G}=
V \frac{3F \left(H^2+K/a^2\right)}{8\pi G}=V\left(\hat{\rho}_{\mathrm{d}}
+\rho_{\mathrm{f}}\right)\,,
\label{eq:2.29}
\end{eqnarray}
where $V=4\pi \tilde{r}_A^3/3$ is the volume inside 
the apparent horizon. The last equality in Eq.~(\ref{eq:2.29}) 
means that $\hat{E}$ corresponds to the total intrinsic energy. 
It is clear from Eq.~(\ref{eq:2.29}) that $\hat{E} >0$ and therefore 
$F>0$, which is consistent with the fact that the effective gravitational 
coupling in $f(R)$ gravity $G_\mathrm{eff} = G/F$ should be 
positive~\cite{Sotiriou:2008rp}. In other words, 
the graviton is not a ghost in the sense of quantum 
theory~\cite{Starobinsky:2007hu}. 
Using Eqs.~(\ref{eq:2.7}) and (\ref{eq:2.17}), we get
\begin{equation}
d\hat{E} = -4\pi \tilde{r}_A^3 H \left(\hat{\rho}_{\mathrm{d}}
+\hat{P}_{\mathrm{d}}+\rho_{\mathrm{f}}+P_{\mathrm{f}}\right) dt 
+4\pi \tilde{r}_A^2 \left(\hat{\rho}_{\mathrm{d}}+\rho_{\mathrm{f}}\right) 
d\tilde{r}_A+\frac{\tilde{r}_A }{2G} dF\,.
\label{eq:2.30}
\end{equation}

It follows from Eqs.~(\ref{eq:2.27}) and (\ref{eq:2.30}) that 
\begin{equation}
T d\hat{S} = d\hat{E}+2\pi \tilde{r}_A^2 
\left(\hat{\rho}_{\mathrm{d}}+\rho_{\mathrm{f}}-\hat{P}_{\mathrm{d}}
-P_{\mathrm{f}}\right) d\tilde{r}_A 
+\frac{\tilde{r}_A}{2G} \left( 1+2\pi \tilde{r}_A T \right) dF\,.
\label{eq:2.31}
\end{equation}
By introducing the work density~\cite{Hayward:1997jp, Hayward:1998ee, 
Cai:2006rs} 
\begin{eqnarray}
\hat{W} \Eqn{\equiv} 
-\frac{1}{2} \left( T^{(\mathrm{matter})\alpha\beta}
h_{\alpha\beta} + \hat{T}^{(\mathrm{d})\alpha\beta} h_{\alpha\beta} 
\right) 
\label{eq:2.32} \\
\Eqn{=} 
\frac12 \left(\hat{\rho}_{\mathrm{d}}+\rho_{\mathrm{f}}
-\hat{P}_{\mathrm{d}}-P_{\mathrm{f}}\right)\,, 
\label{eq:2.33}
\end{eqnarray}
Eq.~(\ref{eq:2.31}) is reduced to
\begin{eqnarray}
T d\hat{S}=-d\hat{E}+\hat{W} dV
+\frac{\tilde{r}_A}{2G} 
\left( 1+2\pi \tilde{r}_A T \right) dF\,, 
\label{eq:2.34}
\end{eqnarray}
which can be rewritten in the form 
\begin{eqnarray}
T d\hat{S}+T d_{i}\hat{S}=-d\hat{E}+\hat{W} dV\,,
\label{eq:2.35}
\end{eqnarray}
where 
\begin{eqnarray}
d_{i}\hat{S} \Eqn{=} -\frac{1}{T} \frac{\tilde{r}_A}{2G}
\left( 1+2\pi \tilde{r}_A T \right) dF
=-\left( \frac{\hat{E}}{T}+\hat{S} \right) \frac{dF}{F} \nonumber \\
\Eqn{=} -\frac{\pi}{G} \frac{4H^2+\dot{H}+3K/a^2}
{\left(H^2+K/a^2\right)\left(2H^2+\dot{H}+K/a^2\right)} dF\,.
\label{eq:2.36}
\end{eqnarray}
Equation (\ref{eq:2.36}) agrees with the result of Ref.~\cite{Wu:2008ir} for 
$K=0$ obtained in $f(R)$ gravity. 

The new term $d_{i}\hat{S}$ can be interpreted as a term of 
entropy produced in the non-equilibrium thermodynamics. 
General relativity with $F={\rm constant}$ leads to $d_{i}\hat{S}=0$, 
which implies that the first-law of equilibrium 
thermodynamics holds. On the other hand, in $f(R)$ gravity 
the additional term in Eq.~(\ref{eq:2.36}) appears because $d F \neq 0$. 

\subsection{Second law of thermodynamics}

To investigate the second law of thermodynamics in $f(R)$ gravity, 
we start with the Gibbs equation in terms of all matter and energy fluid, 
given by 
\begin{eqnarray}
T d\hat{S}_{\mathrm{t}} = d\left( \hat{\rho}_{\mathrm{t}} V \right) +\hat{P}_{\mathrm{t}} dV 
= V d\hat{\rho}_{\mathrm{t}} + \left( \hat{\rho}_{\mathrm{t}} + \hat{P}_{\mathrm{t}} \right) dV\,, 
\label{eq:2.37}
\end{eqnarray}
where $T$ and $\hat{S}_{\mathrm{t}}$ denote the temperature and 
entropy of total energy inside the horizon, respectively. 
Here, we have assumed the same temperature between the outside 
and inside of the apparent horizon. 
To obey the second law of thermodynamics in $f(R)$ gravity, 
we require that~\cite{Wu:2008ir}
\begin{eqnarray}
\frac{d\hat{S}}{dt} + \frac{d\left( d_i \hat{S} \right)}{dt} 
+ \frac{d\hat{S}_{\mathrm{t}}}{dt} \geq 0\,,
\label{eq:2.38}
\end{eqnarray}
which leads to the condition~\cite{Bamba:2009ay} 
\begin{eqnarray}
J \equiv 
\left( \dot{H}-\frac{K}{a^2} \right)^2 F\geq 0\,. 
\label{eq:2.39}
\end{eqnarray}
%
In the FLRW background~(\ref{eq:2.4}), the effective EoS 
$w_\mathrm{eff}$ is given by~\cite{Nojiri:2006ri} 
$w_\mathrm{eff} 
= -1 -2\dot{H}/\left(3H^2\right)$. 
For $\dot{H} < 0$, $w_\mathrm{eff} >-1$, which corresponds to the non-phantom 
(quintessence) phase, 
while for $\dot{H} > 0$, $w_\mathrm{eff} <-1$, which corresponds to the 
phantom phase. 
It follows from Eq.~(\ref{eq:2.39}) that $J \geq 0$ even in the 
phantom phase. 
As a consequence, the second law of thermodynamics in $f(R)$ gravity 
can be satisfied in both phantom and non-phantom phases. 
This consequence is compatible with a phantom model with ordinary 
thermodynamics proposed in Ref.~\cite{Nojiri:2005sr}.

\section{Equilibrium description of thermodynamics in $f(R)$ gravity 
in the Palatini formalism}

The reason why there exists 
a non-equilibrium entropy production term $d_i \hat{S}$ 
is that $\hat{\rho}_{\mathrm{d}}$ and $\hat{P}_{\mathrm{d}}$ defined in 
Eqs.~(\ref{eq:2.10}) and (\ref{eq:2.11}) obey Eq.~(\ref{eq:2.17}), 
whose r.h.s. does not vanish in $f(R)$ gravity because $\dot{F} \neq 0$. 
In other words, the standard continuity equation does not hold. 
In this section, 
we redefine the energy density and pressure of dark components to satisfy the 
continuity equation so that there is no extra entropy production term, 
which is referred as the equilibrium description. 
We will discuss such equilibrium description of thermodynamics in $f(R)$ 
gravity in the Palatini formalism.

\subsection{First law of thermodynamics in equilibrium description}

We rewrite Eqs.~(\ref{eq:2.5}) and (\ref{eq:2.6}) in the following forms: 
\begin{eqnarray}
3F_0 \left( H^2+\frac{K}{a^2} \right) \Eqn{=} 
\kappa^2 \left( \rho_{\mathrm{d}}+\rho_{\mathrm{f}} \right)\,,
\label{eq:3.1} \\
-2F_0 \left( \dot{H}-\frac{K}{a^2} \right)
\Eqn{=} \kappa^2 \left( \rho_{\mathrm{d}}+P_{\mathrm{d}}+\rho_{\mathrm{f}}+P_{\mathrm{f}} \right),
\label{eq:3.2}
\end{eqnarray} 
where $F_0$ is some constant, and 
$\rho_{\mathrm{d}}$ and $P_{\mathrm{d}}$ are 
the energy density and pressure of dark components redefined as 
\begin{eqnarray}
\rho_{\mathrm{d}} \Eqn{\equiv} \frac{1}{\kappa^2} \left[ 
\frac{1}{2} \left( F\mathcal{R} - f \right) 
-3H \dot{F} 
+3\left(F_0-F\right) \left(H^2+\frac{K}{a^2}\right) 
-\frac{3}{4}\frac{\dot{F}^2}{F} 
\right]\,,
\label{eq:3.3}
\\
P_{\mathrm{d}} \Eqn{\equiv} \frac{1}{\kappa^2} 
\left[ 
-\frac{1}{2} \left( F\mathcal{R} - f \right) 
+\ddot{F}+2H \dot{F}
-\left(F_0-F\right) \left(2\dot{H}+3H^2+\frac{K}{a^2}\right) 
-\frac{3}{4}\frac{\dot{F}^2}{F} 
\right]\,.
\label{eq:3.4}
\end{eqnarray} 
It follows from Eqs.~(\ref{eq:3.3}) and (\ref{eq:3.4}) that 
the standard continuity equation can be satisfied
\begin{eqnarray}
\dot{\rho}_{\mathrm{d}}+3H \left(\rho_{\mathrm{d}}+P_{\mathrm{d}}\right)=0\,. 
\label{eq:3.5}
\end{eqnarray}
In this representation, Eq.~(\ref{eq:2.20}) becomes 
\begin{eqnarray}
\frac{F_0}{4\pi G} d\tilde{r}_A=\tilde{r}_A^3 H
\left( \rho_{\mathrm{d}}+P_{\mathrm{d}}+\rho_{\mathrm{f}}+P_{\mathrm{f}} 
\right) dt\,.
\label{eq:3.6}
\end{eqnarray}

By introducing the horizon entropy $S$ in the form 
\begin{eqnarray}
S=\frac{F_0A}{4G}\,,
\label{eq:3.7}
\end{eqnarray}
and using Eq.~(\ref{eq:3.6}), we obtain 
\begin{eqnarray}
\frac{1}{2\pi \tilde{r}_A} dS=4\pi \tilde{r}_A^3 H
\left( \rho_{\mathrm{d}}+P_{\mathrm{d}}+\rho_{\mathrm{f}}+P_{\mathrm{f}} 
\right) dt \,.
\label{eq:3.8}
\end{eqnarray}
{}From the horizon temperature in Eq.~(\ref{eq:2.26}) and Eq.~(\ref{eq:3.8}), 
we find 
\begin{eqnarray}
T dS = 4\pi \tilde{r}_A^3 H \left(\rho_{\mathrm{d}}+P_{\mathrm{d}}+\rho_{\mathrm{f}}+P_{\mathrm{f}} \right) dt 
-2\pi  \tilde{r}_A^2 \left(\rho_{\mathrm{d}}+P_{\mathrm{d}}+\rho_{\mathrm{f}}+P_{\mathrm{f}} \right) d\tilde{r}_A\,.
\label{eq:3.9}
\end{eqnarray}

By defining the Misner-Sharp energy as 
\begin{eqnarray}
E=\frac{F_0 \tilde{r}_A}{2G}=
V\left(\rho_{\mathrm{d}}+\rho_{\mathrm{f}}\right)\,,
\label{eq:3.10}
\end{eqnarray}
we get 
\begin{eqnarray}
dE=-4\pi \tilde{r}_A^3 H \left(\rho_{\mathrm{d}}+P_{\mathrm{d}}+\rho_{\mathrm{f}}+P_{\mathrm{f}}\right) dt 
+4\pi \tilde{r}_A^2 \left(\rho_{\mathrm{d}}+\rho_{\mathrm{f}}\right) d\tilde{r}_A\,,
\label{eq:3.11}
\end{eqnarray}
where there does not exist any additional term proportional to $dF$ 
on the r.h.s. due to the continuity equation (\ref{eq:3.5}). 
By combining Eqs.~(\ref{eq:3.9}) and (\ref{eq:3.11}), we obtain 
the following equation corresponding to the first law of 
equilibrium thermodynamics: 
\begin{eqnarray}
T dS=-dE+W dV\,,
\label{eq:3.12}
\end{eqnarray}
where the work density $W$ is defined by 
\begin{eqnarray}
W=\frac12 \left( \rho_{\mathrm{d}}+\rho_{\mathrm{f}}-P_{\mathrm{d}}-P_{\mathrm{f}} \right)\,.
\label{eq:3.13}
\end{eqnarray}
As a result, 
an equilibrium description of thermodynamics can be derived 
by redefining the energy density $\rho_{\mathrm{d}}$ and 
the pressure $P_{\mathrm{d}}$ so that the continuity equation (\ref{eq:3.5}) 
can be met. 

Note that the constant $F_0$ can be chosen arbitrarily 
as long as $F_0>0$ in order to ensure that the sign of the 
Friedmann equation (\ref{eq:3.1}) does not change. 
It is considered that the natural choice is $F_0=1$ because in this case 
the entropy $S$ and the Misner-Sharp energy $E$ reduce to the standard forms 
in the Einstein gravity: $S=A/\left(4G\right)$ and 
$E=\tilde{r}_A/\left(2G\right)$, respectively.

{}From Eqs.~(\ref{eq:3.1}), (\ref{eq:3.2}) and (\ref{eq:3.8}), we obtain 
\begin{eqnarray}
\dot{S} = 
8\pi^2 H \tilde{r}_A^4 \left(\rho_{\mathrm{d}}+\rho_{\mathrm{f}}+P_d+
P_{\mathrm{f}}\right) 
= -\frac{2\pi F_0}{G} \frac{H \left(\dot{H}-K/a^2\right)}
{\left(H^2+K/a^2\right)^2}\,.
\label{eq:3.14}
\end{eqnarray}
Thus, 
the horizon entropy increases as long as the null energy condition 
$\rho_{\mathrm{t}}+P_{\mathrm{t}} \equiv \rho_{\mathrm{d}}+\rho_{\mathrm{f}}
+P_{\mathrm{d}}+P_{\mathrm{f}} \ge 0$ is satisfied. 

The reasons why the above equilibrium picture of thermodynamics can be 
realized are as follows: 
The first is that there exists an energy momentum tensor 
$T_{\mu \nu}^{({\mathrm{d}})}$ satisfying the local conservation law
$\nabla^{\mu} T_{\mu \nu}^{({\mathrm{d}})}=0$. 
The second is that the entropy $S$ is given by Eq.~(\ref{eq:3.7}) 
as in general relativity, 
corresponding to the Einstein equation in the form 
\begin{eqnarray}
G_{\mu \nu}=\frac{8\pi G}{F_0} 
\left(  T_{\mu \nu}^{(\rm matter)}
+T_{\mu \nu}^{({\mathrm{d}})} \right)\,,
\label{eq:3.15}
\end{eqnarray}
where 
\begin{eqnarray}
T_{\mu \nu}^{({\mathrm{d}})} 
\Eqn{\equiv} 
\frac{1}{\kappa^2}
\biggl\{ 
-\frac{1}{2} g_{\mu \nu} \left[ F\left(\mathcal{R}-R\right) + F_0 R - f 
\right] 
+\nabla_{\mu}\nabla_{\nu}F -g_{\mu \nu} \Box F 
+\left( F_0-F \right)R_{\mu \nu} 
\nonumber \\
&&
{}
-\frac{3}{2F} \left[ \nabla_{\mu} F \nabla_{\nu}F 
-\frac{1}{2} g_{\mu \nu} \left( \nabla F \right)^2 \right]
\biggr\}\,.
\label{eq:3.16}
\end{eqnarray}
It is clear that 
the local conservation of $T_{\mu \nu}^{({\mathrm{d}})}$ follows from 
Eq.~(\ref{eq:3.15}) due to the relations 
$\nabla^{\mu}G_{\mu \nu}=0$ and
$\nabla^{\mu}T_{\mu \nu}^{(\rm matter)}=0$.
By using the conserved energy momentum tensor $T_{\mu \nu}^{({\mathrm{d}})}$
and the horizon entropy $S$ defined in Eq.~(\ref{eq:3.7}) and 
following the method in Ref.~\cite{Jacobson:1995ab}, 
the Einstein equation (\ref{eq:3.15}) can also be derived. 

It can be shown that the horizon entropy $S$ in the equilibrium 
description has the following relation with $\hat{S}$ in the 
non-equilibrium description~\cite{Bamba:2009id}:
\begin{equation}
dS = d\hat{S} + d_i \hat{S} 
+\frac{\tilde{r}_A}{2GT} dF 
-\frac{2\pi \left(F_0-F\right)}{G}
\frac{H \left(\dot{H}-K/a^2\right)}{\left(H^2+K/a^2\right)^2}\, dt.
\label{eq:3.A-1}
\end{equation}
By using the relations (\ref{eq:2.36}) and (\ref{eq:3.8}), 
Eq.~(\ref{eq:3.A-1}) is rewritten to the following form: 
\begin{equation}
dS=\frac{F_0}{F} d\hat{S}+\frac{F_0}{F}
\frac{2H^2+\dot{H}+K/a^2}{4H^2+\dot{H}+3K/a^2}\,d_i \hat{S}\,,
\label{eq:3.A-2}
\end{equation}
where 
\begin{equation}
d_i \hat{S}=-\frac{6\pi}{G} \frac{4H^2+\dot{H}+3K/a^2}
{H^2+K/a^2} \frac{dF}{R}\,.
\label{eq:3.A-3}
\end{equation}
The difference between $S$ and $\hat{S}$ appears in $f(R)$ gravity due to 
$dF \neq 0$, although 
$S$ is identical to $\hat{S}$ in general relativity ($F=F_0=1$). 
{}From Eq.~(\ref{eq:3.A-2}), we see that the change of the horizon entropy $S$ 
in the equilibrium framework involves the information of 
both $d\hat{S}$ and $d_i \hat{S}$ in the non-equilibrium framework.

In the flat FLRW spacetime ($K=0$), 
the Bekenstein-Hawking entropy is simply proportional to the inverse squared 
of the expansion rate of the universe ($S \propto H^{-2}$)
independent of gravitational theories. 
Hence $S$ grows as long as $H$ decreases, whereas
the increase of $H$ leads to the decrease of 
$S$ (as in the case of superinflation). 
In other words, this property mimics the standard general relativistic 
picture with energy density $\rho_{\mathrm{d}}$ and pressure $P_{\mathrm{d}}$ 
of dark components defined in Eqs.~(\ref{eq:3.3}) and (\ref{eq:3.4}). 
It can be clearly understood that the Bekenstein-Hawking entropy grows for 
$w_{\mathrm{t}}=P_{\mathrm{t}}/\rho_{\mathrm{t}} > -1$, 
where $\rho_{\mathrm{t}} \equiv \rho_{\mathrm{d}}+\rho_{\mathrm{f}}$ and 
$P_{\mathrm{t}} \equiv P_{\mathrm{d}}+P_{\mathrm{f}}$, 
and that it decreases for $w_{\mathrm{t}}<-1$.

The Wald entropy carries the information of gravitational theories 
through the dependence $\hat{S} \propto FH^{-2}$. 
For example, 
in a model of $f(R)$ gravity $f(R)= R + \alpha R^n$, where 
$\alpha$ and $n$ are constants, 
it follows that $\hat{S} \propto H^{2(n-2)}$ and hence 
$\hat{S}$ grows apart from $n=2$ 
because $H$ increases (decreases) for $n>2$ ($n<2$). 
This evolution is different from that of the Bekenstein-Hawking entropy. 
The introduction of the entropy production term $d_{i}\hat{S}$ 
in the non-equilibrium framework allows us to have a connection 
with the equilibrium picture based on the Bekenstein-Hawking entropy 
as given in Eq.~(\ref{eq:3.A-1}). 
Thus, it is considered that our equilibrium description of thermodynamics 
is useful not only to provide the general relativistic analogue 
of the horizon entropy irrespective of gravitational theories 
but also to understand the non-equilibrium thermodynamics 
deeper in connection with the standard equilibrium framework. 

Finally, we remark that the EoS of dark components in the 
non-equilibrium description is different from that in the 
equilibrium description. 
It follows from Eqs.~(\ref{eq:2.10}) and (\ref{eq:2.11}) that 
the EoS of dark components in the non-equilibrium description 
$\hat{w}_{\mathrm{d}} = \hat{P}_{\mathrm{d}} / \hat{\rho}_{\mathrm{d}}$ is 
given by 
\begin{equation}
\hat{w}_{\mathrm{d}} 
= \frac{-2 \left( F\mathcal{R} - f \right) 
+4\ddot{F}+8H\dot{F} -3\dot{F}^2/F}
{2 \left( F\mathcal{R} - f \right) 
-12H\dot{F} -3\dot{F}^2/F}\,.
\label{eq:3.A-4}
\end{equation}
On the other hand, 
by using Eqs.~(\ref{eq:3.3}) and (\ref{eq:3.4}), 
the EoS of dark components in the equilibrium description 
$w_{\mathrm{d}} = P_{\mathrm{d}} / \rho_{\mathrm{d}}$ is 
expressed as 
\begin{equation}
w_{\mathrm{d}} 
= \frac{-2\left( F\mathcal{R} - f \right) 
+4\ddot{F}+8H \dot{F}
-4\left(F_0-F\right) \left(2\dot{H}+3H^2+K/a^2\right) 
-3\dot{F}^2/F}
{2\left( F\mathcal{R} - f \right) 
-12H \dot{F} 
+12\left(F_0-F\right) \left(H^2+K/a^2\right) 
-3\dot{F}^2/F}\,. 
\label{eq:3.A-5}
\end{equation}
It is clear from Eqs.~(\ref{eq:3.A-4}) and (\ref{eq:3.A-5}) that in general 
$\hat{w}_{\mathrm{d}} \neq w_{\mathrm{d}}$ except general relativity 
with $F=F_0={\rm constant}$. 
Thus, in order to compare the EoS of dark components predicted by a theory of 
$f(R)$ gravity with the observations, it is necessary to derive the expression 
of the EoS of dark components in both non-equilibrium description and 
equilibrium one. This is another important physical motivation to 
consider both non-equilibrium and equilibrium descriptions. 

\subsection{Second law of thermodynamics in equilibrium description}

To examine the second law of thermodynamics in the equilibrium description, 
we write the Gibbs equation in terms of all matter and energy fluid as 
\begin{eqnarray}
T dS_{\mathrm{t}} = d\left( \rho_{\mathrm{t}} V \right) + P_{\mathrm{t}} dV 
= V d\rho_{\mathrm{t}} + \left( \rho_{\mathrm{t}} + P_{\mathrm{t}} \right) 
dV\,. 
\label{eq:3.17}
\end{eqnarray}

The second law of thermodynamics can be described by 
\begin{eqnarray}
\frac{dS}{dt} + \frac{dS_{\mathrm{t}}}{dt} \geq 0\,, 
\label{eq:3.18}
\end{eqnarray}
which gives 
\begin{eqnarray}
\frac{12\pi F_0}{G} \frac{H \left( \dot{H}-K/a^2 \right)^2}
{\left( H^2+K/a^2 \right)^2}\frac{1}{R} \geq 0\,
\label{eq:3.19}
\end{eqnarray}
by using $V=4\pi \tilde{r}_A^3/3$, 
$R=6 \left( \dot{H} + 2H^2+K/a^2 \right)$, and 
Eqs.~(\ref{eq:2.26}), (\ref{eq:3.2}) and (\ref{eq:3.14}). 
It is clear that 
in the flat FLRW spacetime ($K=0$), 
the second law of thermodynamics can be met in both non-phantom and 
phantom phases, which is the same as the non-equilibrium description. 
This also agrees with the argument in Refs.~\cite{Gong:2006ma, Jamil:2009eb}. 
We remark that the above consequence can be shown explicitly only for the same 
temperature of the universe outside and inside the apparent horizon. 
As a result, a unified understanding between non-equilibrium 
and equilibrium pictures of thermodynamics has been obtained.

\section{Conclusion}

In the present paper, we have studied the first and second laws of 
thermodynamics of the apparent horizon in $f(R)$ gravity 
in the Palatini formalism. 
We have explored both non-equilibrium and equilibrium descriptions of 
thermodynamics in $f(R)$ gravity. 
The equilibrium framework is more transparent than 
the non-equilibrium one because in the equilibrium description the horizon 
entropy is described by the single expression $S$, 
whereas in the non-equilibrium description 
it consists of two contributions, 
the change of the horizon entropy $d\hat{S}$ and the term 
produced in the non-equilibrium thermodynamics $d_{i}\hat{S}$. 
We have also shown that the second law of 
thermodynamics can be satisfied in not only the non-phantom phase but also 
the phantom one, provided that the temperature of the universe inside the 
horizon is equal to that of the apparent horizon.

\section*{Acknowledgments}

K.B. acknowledges the KEK theory exchange program 
for physicists in Taiwan and the very kind hospitality at 
KEK and Tokyo University of Science. 
The work is supported in part by 
the National Science Council of R.O.C. under: 
Grant \#s: NSC-95-2112-M-007-059-MY3 and
NSC-98-2112-M-007-008-MY3
and 
National Tsing Hua University under the Boost Program and Grant \#: 
97N2309F1.


\end{document}